# Extending the Low-Frequency Limit of Time-Domain Thermoreflectance via Periodic Waveform Analysis

Mingzhen Zhang[1], Tao Chen[1], Shangzhi Song[1], Yunjia Bao[2,3], Ruiqiang Guo[3], Weidong Zheng[3,*], Puqing Jiang[1,*], Ronggui Yang[1,4,*]

[1] School of Energy and Power Engineering, Huazhong University of Science and Technology, Wuhan, Hubei 430074, China
[2] Institute of Novel Semiconductors, State Key Laboratory of Crystal Materials, Shandong University, Jinan, Shandong 250100, China
[3] Thermal Science Research Center, Shandong Institute of Advanced Technology, Jinan, Shandong 250103, China
[4] Department of Energy and Resource Engineering, College of Engineering, Peking University, Beijing 100871, China

**ABSTRACT:** Time-domain thermoreflectance (TDTR) is a powerful technique for characterizing the thermal properties of layered materials. However, its effectiveness at modulation frequencies below 0.1 MHz is hindered by pulse accumulation effects, limiting its ability to accurately measure in-plane thermal conductivities below 6 W/(m · K). Here, we present a periodic waveform analysis-based TDTR (PWA-TDTR) method that extends the measurable frequency range down to 50 Hz with minimal modifications to the conventional setup. This advancement greatly enhances measurement sensitivity, enabling accurate measurements of in-plane thermal conductivities as low as 0.2 W/(m · K). We validate the technique by measuring polymethyl methacrylate (PMMA) and fused silica, using PWA-TDTR to obtain in-plane thermal diffusivity and conventional TDTR to measure cross-plane thermal effusivity. Together, these allow the extraction of both thermal conductivity and volumetric heat capacity, with results in excellent agreement with literature values. We further demonstrate the versatility of PWA-TDTR through (1) thermal conductivity and heat capacity measurements of thin liquid films and (2) depth-resolved thermal conductivity profiling in lithium niobate crystals, revealing point defect-induced inhomogeneities at depths up to 100 μm. By overcoming frequency and sensitivity constraints, PWA-TDTR significantly expands the applicability of TDTR, enabling detailed investigations of thermal transport in materials and conditions that were previously challenging to study.

*Corresponding Authors: weidong.zheng@iat.cn; jpq2021@hust.edu.cn; ronggui@pku.edu.cn

## I. INTRODUCTION

Time-domain thermoreflectance (TDTR) is a powerful and widely used technique for characterizing the thermal properties of bulk and thin-film materials [1-5]. In this method, a modulated train of femtosecond laser pulses is used to periodically heat the sample, while a phase-locked probe pulse train monitors changes in reflectance at precisely controlled time delays. With its picosecond temporal resolution, TDTR enables accurate measurement of thermal conductivity, volumetric heat capacity, and interfacial thermal conductance. These capabilities have made TDTR an indispensable tool in the development of advanced thermal management solutions for electronics, photonics, and other technologies where precise thermal property data are essential.

Despite its broad applicability, TDTR is constrained by a typical modulation frequency range of 0.1 to 20 MHz [6]. At frequencies below 0.1 MHz, pulse accumulation effects become significant, leading to substantial uncertainty in determining the reference phase and compromising the accuracy of thermal property measurements. This limitation restricts the ability to study materials that require low modulation frequencies for adequate thermal penetration depths, leaving a notable gap in the capabilities of conventional TDTR. For example, TDTR faces challenges in accurately measuring in-plane thermal conductivities below 6 W/(m · K) [7].

Efforts have been made to address these challenges. For instance, Larkin et al. [8] proposed a variant known as "low-rep" TDTR, which employs laser pulses with a reduced repetition rate of 250 kHz instead of the conventional 80 MHz. The lower repetition rate helps mitigate pulse accumulation effects, allowing modulation frequencies to be extended below 0.1 MHz. However, "low-rep" TDTR has challenges in both instrumentation and thermal modeling. The reduced repetition rate increases experimental complexity, making it difficult to switch between "low-rep" and conventional TDTR within the same setup. Moreover, it requires the inclusion of many frequency-domain terms in the thermal model, significantly increasing computation time. These drawbacks limit the practical adoption of the "low-rep" TDTR approach. Therefore, there is still a pressing need for robust solutions to extend the low-frequency limit of TDTR while maintaining accuracy and ease of operation.

Recently, the square-pulsed source (SPS) method [9], a thermoreflectance-based technique, has emerged as a promising solution for low-frequency thermal measurements. This approach utilizes a square-wave-modulated continuous-wave laser as the heating source, while a second continuous-wave laser monitors temperature-induced reflectance changes. By applying periodic waveform analysis (PWA), the SPS method captures time-resolved amplitude signals and supports a broad modulation frequency range from 1 Hz to 10 MHz. This extended modulation frequency range enables accurate measurement of in-plane thermal conductivities as low as $0.2\ \mathrm{W/(m \cdot K)}$ [9].

In this study, we adapt the PWA approach to extend the low-frequency capability of TDTR, developing a technique we term "PWA-TDTR". Instead of mitigating the pulse accumulation effect, PWA-TDTR leverages it at low modulation frequencies, effectively treating the 80-MHz laser pulse train as a continuous-wave heat source. This method integrates seamlessly with existing TDTR setups, requiring only minimal modifications and enabling rapid switching between conventional TDTR and PWA-TDTR modes. With modulation frequencies extended down to 50 Hz, PWA-TDTR offers significantly enhanced sensitivity and accuracy, providing a practical and flexible solution to overcome the low-frequency limitations of traditional TDTR.

We validate the accuracy of PWA-TDTR by measuring the in-plane thermal diffusivities of polymethyl methacrylate (PMMA) and fused silica. By combining these measurements with cross-plane thermal effusivities obtained using conventional TDTR, we accurately determine both the thermal conductivity and volumetric heat capacity of each material. To further demonstrate the versatility of PWA-TDTR, we apply it to (1) simultaneously measure the thermal conductivity and heat capacity of thin liquid films, and (2) resolve depth-dependent thermal conductivity variations in lithium niobate crystals, revealing inhomogeneities up to 100 μm beneath the surface. These results highlight the robustness and broad applicability of PWA-TDTR, effectively addressing a key limitation of conventional TDTR in low-frequency thermal measurements.

## II. METHODOLOGIES

### A. Principle and implementation of PWA-TDTR

Figure 1 shows the schematic of the PWA-TDTR setup. While preserving all the core functionalities of a conventional TDTR system, PWA-TDTR incorporates key enhancements to enable low-frequency measurements. The primary modifications include a balanced photodetector (PDA465A, Thorlabs) and a Zurich Instruments UHF lock-in amplifier equipped with an upgraded periodic waveform analysis (PWA) module.

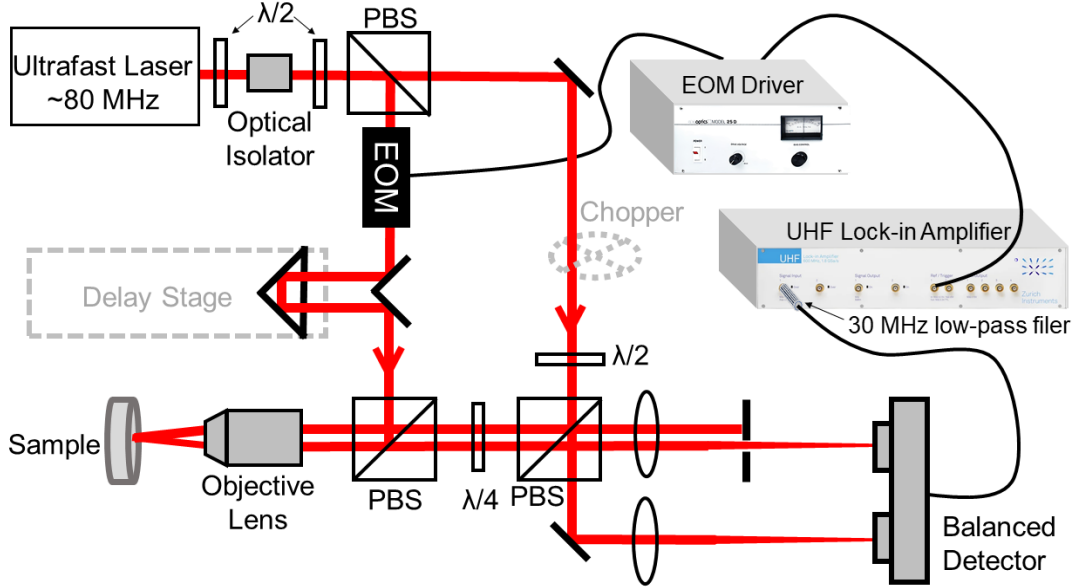

**FIG. 1** Schematic of the PWA-TDTR setup. (PBS: Polarizing beam splitter; EOM: Electro-optic modulator; $\lambda/2$: Half-wave plate; $\lambda/4$: Quarter-wave plate.)

During low-frequency measurements in PWA-TDTR, the UHF lock-in amplifier generates a 50% duty cycle square wave to drive the electro-optic modulator (EOM), while simultaneously providing an internal reference signal for the lock-in detection. Unlike conventional TDTR, which varies the delay time, PWA-TDTR operates at a fixed negative delay (e.g., -20 ps), and the optical chopper in the probe path is deactivated. The PWA module of the UHF lock-in amplifier captures time-resolved amplitude signals across the entire square-wave cycle. These signals are subsequently analyzed using a heat transfer model to extract thermal properties.

To enable meaningful comparison between experimental and simulated signals, the captured signals undergo normalization in both amplitude and time. Specifically, the time axis is normalized by the heating period. For amplitude normalization, the signal is scaled such that it takes values of 0 and 1 at the normalized times of -0.02 and 0.48, respectively. While these reference points are arbitrarily chosen, any consistent normalization scheme may be used, as long as it is applied identically to both experimental and simulated data.

## B. Thermal model for PWA-TDTR

The thermal model used in PWA-TDTR shares the same physical foundation as that of conventional TDTR, with the primary distintion lying in the mathematical treatment of harmonic components. Both methods employ an identical experimental configuration, in which an 80 MHz pulse train is modulated by a square wave and focused onto the sample as a Gaussian-distributed heat source. However, the two approaches differ in how they handle the frequency components of the modulation.

In conventional TDTR, the heat source is modeled as[10]:

$$p_1^{\text{TDTR}}(r,t) = \frac{2A_1}{\pi r_1^2} \exp\left(-\frac{2r^2}{r_1^2}\right) e^{i\omega_0 t} \sum_{n=-\infty}^{\infty} \delta(t - nT_s - t_0) \quad (1)$$

In contrast, the heat source in PWA-TDTR is described by:

$$p_1^{\text{PWA-TDTR}}(r,t) = \frac{2A_1}{\pi r_1^2} \exp\left(-\frac{2r^2}{r_1^2}\right) \sum_{m=0}^{\infty} \frac{4}{i(2m+1)\pi} e^{i(2m+1)\omega_0 t} \sum_{n=-\infty}^{\infty} \delta(t - nT_s - t_0) \quad (2)$$

Here, $A_1$ denotes the average pump power, spatially distributed as a Gaussian beam with a $1/e^2$ radius $r_1$. Temporally, the pump consists of an 80 MHz pulse train (with period $T_s = 1/f_{rep}$), time-shifted by $t_0$, and modulated by a square wave at angular frequency $\omega_0$. In conventional TDTR, only the fundamental modulation component $e^{i\omega_0 t}$ is considered. By contrast, PWA-TDTR retains the full

Fourier series of the square wave, incorporating all odd harmonics $(2m + 1)\omega_0$, each scaled by the coefficient $\frac{4}{i(2m+1)\pi}$.

Through detailed derivations (see Supplementary Information, Section S1), the time-domain temperature response signal measured at a fixed delay time $t_d$ in PWA-TDTR is given by:

$$\Delta\Theta^{\text{PWA-TDTR}}(t) = \sum_{m=0}^{\infty} \frac{4}{i(2m+1)\pi} e^{i(2m+1)\omega_0 t} \frac{\omega_s}{2\pi} \sum_{n=-\infty}^{\infty} \Delta T(n\omega_s + (2m+1)\omega_0) e^{in\omega_s t_d} \quad (3)$$

In this expression, $\Delta T(\omega)$ represents the frequency-domain temperature response of the thermal system to harmonic heating at frequency $\omega$, which is expressed as:

$$\Delta T(\omega) = A_1 A_2 \int_0^{\infty} \hat{G}(\rho, \omega) \exp(-\pi^2 \rho^2 r_0^2) \, 2\pi \rho \, d\rho \quad (4)$$

Here, $A_2$ is the average power of the probe beam, which is also Gaussian-distributed in space with a $1/e^2$ radius $r_2$. The root mean square (RMS) average of the pump and probe sizes is defined as $r_0 = \sqrt{(r_1^2 + r_2^2)/2}$. $\hat{G}(\rho, \omega)$ is the Green's function in the frequency domain, describing the temperature response of the multilayer system to a unit heat flux at spatial frequency ρ and angular frequency ω. Detailed formulations of the Green's function for multilayer structures in thermoreflectance experiments can be found in prior literature [10].

The expression for $\Delta\Theta^{\text{PWA-TDTR}}(t)$ in Eq. (3) is a complex-valued function, but only its real part corresponds to the physically measurable signal. By extracting the real component and applying appropriate normalization in both amplitude and time, the simulated response can be directly compared with experimental measurements.

For comparison, in conventional TDTR, the time-domain temperature response at a fixed delay time $t_d$ is given by:

$$\Delta\Theta^{\text{TDTR}}(t) = e^{i\omega_0 t} \frac{\omega_s}{2\pi} \sum_{n=-\infty}^{\infty} \Delta T(n\omega_s + \omega_0) e^{in\omega_s t_d} \quad (5)$$

This signal is sinusoidal, with an amplitude expressed as:

$$\Delta R = \sum_{n=-\infty}^{\infty} \Delta T(n\omega_s + \omega_0) e^{in\omega_s t_d} \quad (6)$$

In conventional TDTR measurements, the real and imaginary parts of $\Delta R$ correspond to the in-phase and out-of-phase signals detected by the lock-in amplifier, respectively.

Comparing Eqs. (3) and (5), it is evident that the PWA-TDTR signal can be interpreted as a superposition of TDTR signals at multiple modulation frequencies, specifically the odd harmonics of the square wave modulation. Each harmonic component is weighted by a factor of $\frac{4}{i(2m+1)\pi}$, consistent with the linear response of the thermal system.

However, simulating a full set of PWA-TDTR signals is computationally demanding: it typically requires ~1000 times longer than conventional TDTR simulations, with a single signal set often taking several minutes on a standard personal computer. This significantly limits its practicality for iterative data fitting.

Fortunately, in measurement conditions where pulse accumulation dominates, such as when using low modulation frequencies, targeting low-conductivity materials, and measuring at a negative delay time, the PWA-TDTR signal closely approximate that of a continuous-wave (CW) laser system. In such cases, the time-domain response simplifies to the expression derived in Ref. [9]:

$$\Delta\Theta^{CW}(t) = \frac{1}{2}\Delta T(0) + \text{Re}\left[\sum_{n=1}^{\infty} \text{sinc}\left(\frac{n}{2}\right) e^{in(\omega_0 t - \frac{\pi}{2})} \Delta T(n\omega_0)\right] \qquad (7)$$

where $\Delta T(\omega)$ is defined in Eq. (4), and Re[$x$] denotes taking the real part of the complex number $x$.

In practical implementations of PWA-TDTR, a negative delay time (e.g., -20 ps) is typically chosen to suppress transient pulsing effects. Additionally, low modulation frequencies are employed to ensure that $\Delta\Theta^{CW}(t)$ provides a good approximation to $\Delta\Theta^{PWA-TDTR}(t)$. This allows the use of the computationally efficient CW model for accurate signal fitting and enhance analysis efficiency.

As with other thermoreflectance techniques, sensitivity analysis plays a critical role in guiding the signal fitting process. In PWA-TDTR, the sensitivity coefficient for a parameter $\xi$ is defined as:

$$S_\xi = \frac{\partial A_{\text{norm}}/A_{\text{norm}}}{\partial \xi/\xi} \qquad (8)$$

where $\xi$ denotes any parameter in the thermal model under analysis, and $A_{\text{norm}}$ is the normalized amplitude of the PWA-TDTR signal. By this definition, $S_\xi$ quantifies how a 1% change in $\xi$ leads to an $S_\xi$% change in $A_{\text{norm}}$.

Sensitivity analysis is also essential for uncertainty quantification. We employ a comprehensive error propagation scheme to estimate the uncertainties in multiple fitted parameters, based on fitting multiple sets of measured signals. This method accounts for uncertainties in all input parameters and includes the statistical deviations between experimental and simulated signals based on the best fitting. Further details regarding the uncertainty estimation procedure are available in references [11-13].

## III. RESULTS AND DISCUSSION

### A. Validation of PWA-TDTR on standard samples

To validate the feasibility of PWA-TDTR, we applied it to measure PMMA and silica samples, with the corresponding signals and data processing presented in Fig. 2.

Figure 2(a1) shows the signals measured on an Al/PMMA sample using PWA-TDTR with a spot size of $r_0 = 9.2$ μm and a modulation frequency of 50 Hz. The square-wave heating induces signals of both heating and cooling phases, as illustrated in the inset. For clarity, only the cooling-phase signals are shown on a logarithmic scale. The best fit is achieved using the CW laser-based thermal model ($\Delta\Theta(t)^{CW}$ as in Eq. (4)), with model predictions plotted as a solid curve. For comparison, predictions from the pulsed laser model ($\Delta\Theta(t)^{PWA-TDTR}$ as in Eq. (2)) are shown as a dashed curve in the same plot, confirming that both models yield identical results.

Figure 2(a2) presents the corresponding sensitivity coefficients for all thermal system parameters. The signals are primarily sensitive to the in-plane thermal conductivity ($k_r$) and heat capacity ($C$) of PMMA, as well as the laser spot size ($r_0$). These parameters are correlated, with their sensitivity coefficients satisfying the relation $S_{r_0} = 2S_C = -2S_{k_r}$, indicating that the signals are predominantly sensitive to the combined parameter $k_r/(Cr_0^2)$ [14, 15]. With $r_0$ pre-determined using the beam-offset method [7] with a 2% uncertainty, the best fit of the signals in Fig. 2(a1) yields an in-plane thermal diffusivity for PMMA of $k_r/C = 0.1156 \times 10^{-6}$ m$^2$/s, with an estimated uncertainty of 4%.

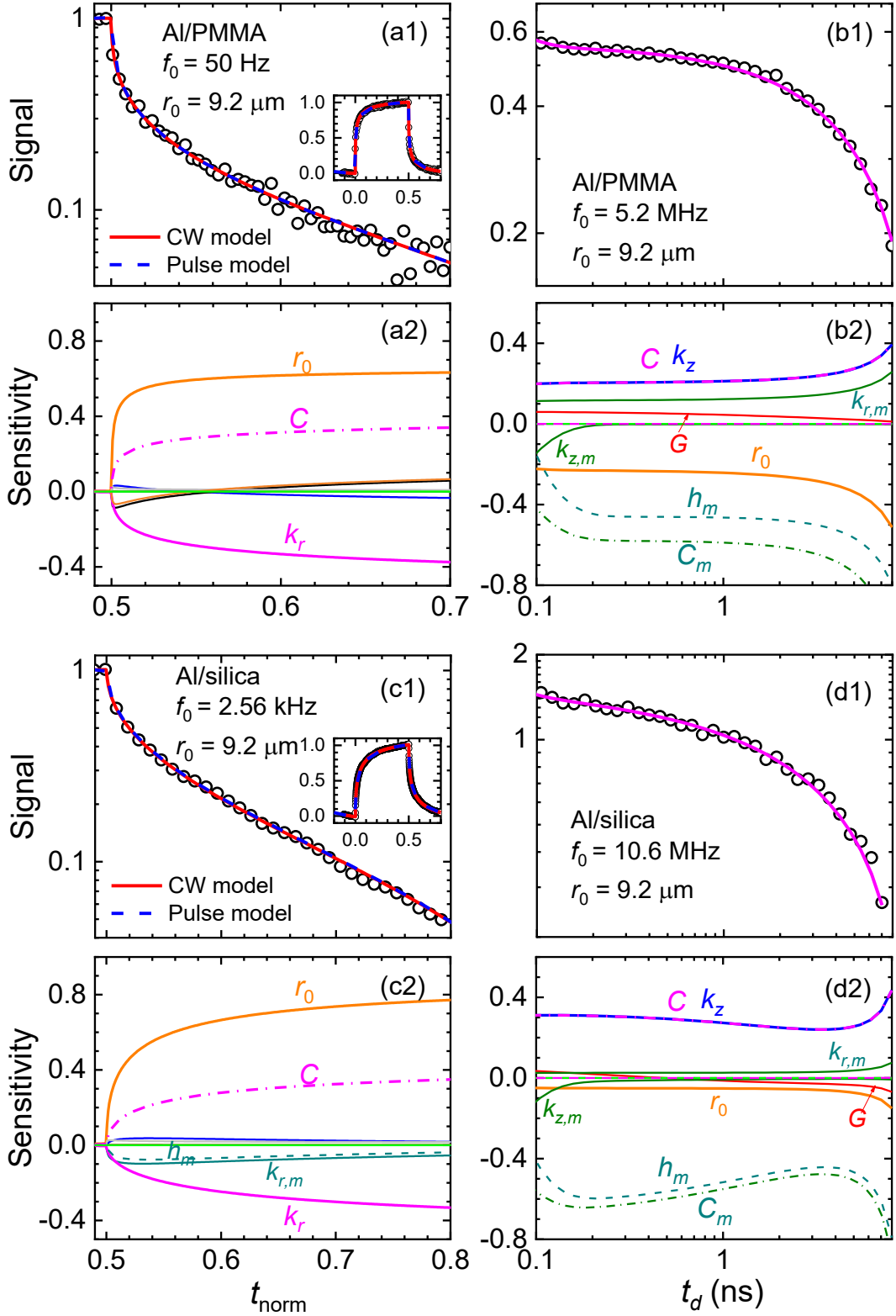

**FIG. 2** Measured signals and sensitivity analysis for PMMA and silica samples using PWA-TDTR and conventional TDTR methods. (a) PMMA measured with PWA-TDTR at 50 Hz; (b) PMMA measured with conventional TDTR at 5.2 MHz; (c) Silica measured with PWA-TDTR at 2.56 kHz; (d) Silica measured with conventional TDTR at 10.6 MHz. (a1-d1) Corresponding measured signals; (a2-d2) Sensitivity analysis.

Figure 2(b1) shows the signals measured on the same sample using the conventional TDTR method with the same spot size at a modulation frequency of 5.2 MHz. Sensitivity analysis in Fig.

2(b2) reveals that this signal set is primarily sensitive to two combined parameters: $\sqrt{k_z C}/(h_m C_m)$ and $k_{r,m}/(C_m r_0^2)$, with additional sensitivity to a third combined parameter: $G/(h_m C_m)$. With the Al film thickness determined using picosecond acoustics as $h_m = 107 \pm 5$ nm, the Al heat capacity from literature [16] as $C_m = 2.44 \pm 0.07$ MJ/(m$^3 \cdot$K), and the in-plane thermal conductivity of the Al film derived from its electrical resistivity via the Wiedemann-Franz law as $k_{r,m} = 110 \pm 11$ W/(m·K), best-fitting this set of signals yields the product of cross-plane thermal conductivity and heat capacity of PMMA as $k_z C = 0.3124 \times 10^6$ W$^2 \cdot$s/(m$^4 \cdot$K$^2$), with an estimated uncertainty of 15%.

Since PMMA is an amorphous polymer with no preferred molecular orientation under standard processing conditions, it can be reasonably approximated as thermally isotropic ($k = k_r = k_z$). By combining the two sets of measurements under this isotropic assumption, the thermal conductivity of PMMA is determined as $k_{\text{PMMA}} = 0.19 \pm 0.015$ W/(m·K), and its heat capacity as $C_{\text{PMMA}} = 1.644 \pm 0.13$ MJ/(m$^3$·K), both with an uncertainty of ~8%. The uncertainties of $k_{\text{PMMA}}$ and $C_{\text{PMMA}}$ are calculated using error propagation: $\eta_k = \eta_C = \frac{1}{2}\sqrt{(\eta_{k_z C})^2 + (\eta_{k_r/C})^2}$. The obtained values for $k_{\text{PMMA}}$ and $C_{\text{PMMA}}$ are in good agreement with established literature data [17, 18].

We further validate the methods on a fused silica sample. Figure 2(c1) shows the signals measured on an Al/silica sample using the PWA-TDTR method with a spot size of $r_0 = 9.2$ μm and a modulation frequency of 2.56 kHz. The signals predicted by both the CW laser model (solid curve) and the pulsed laser model (dashed curve) are identical, confirming that both models can accurately fit this set of signals. Sensitivity analysis in Fig. 2(c2) indicates that the signals are primarily sensitive to the combined parameter $k_r/(C r_0^2)$, with additional sensitivity to $k_{r,m}$ and $h_m$. With $r_0$, $k_{r,m}$, and $h_m$ pre-determined, best-fitting the signals in Fig. 2(c1) yields an in-plane thermal diffusivity for silica of $k_r/C = 0.836 \times 10^{-6}$ m$^2$/s, with an estimated uncertainty of 6%.

Figure 2(d1) shows the signals measured on the same silica sample using the conventional TDTR method with the same spot size at a modulation frequency of 10.6 MHz. Sensitivity analysis in Fig. 2(d2) reveals that this signal set is primarily sensitive to the combined parameter $\sqrt{k_z C}/(h_m C_m)$.

With $h_m$ and $C_m$ pre-determined, best-fitting this set of signals yields the product of cross-plane thermal conductivity and heat capacity of silica as: $k_z C = 2.31 \times 10^6$ W$^2 \cdot$s/(m$^4 \cdot$K$^2$), with an estimated uncertainty of 12%.

As fused silica is an amorphous material with no long-range order, its thermal conductivity can be reasonably assumed to be isotropic with $k = k_r = k_z$. By combining the two sets of measurements under this assumption, the thermal conductivity of silica is determined to be $k_{\text{silica}} = 1.39 \pm 0.09$ W/(m·K), and its heat capacity as $C_{\text{silica}} = 1.65 \pm 0.11$ MJ/(m$^3$·K), both with an uncertainty of ~7%. These values for $k_{\text{silica}}$ and $C_{\text{silica}}$ are in good agreement with established literature values [19, 20].

## B. Simultaneously measuring thermal conductivity and heat capacity of thin liquid films

As an advanced application of PWA-TDTR, we employ this technique to measure the thermal conductivity and heat capacity of thin liquid films. The measurement relies on a transparent substrate coated with a metal transducer layer, a configuration previously adopted for thermal characterization of confined liquids [e.g., Refs. [21-25]]. Building on this established framework, our approach consists of two steps: a calibration measurement on the bare substrate, followed by a formal measurement with the liquid applied. Each step is performed using PWA-TDTR at a low modulation frequency and conventional TDTR at a high frequency, respectively.

Figure 3(a1) presents the signals measured on an Al/glass sample using PWA-TDTR with a spot size of $r_0 = 15$ μm and a modulation frequency of 100 Hz. Sensitivity analysis in Fig. 3(a2) suggests that these signals are primarily sensitive to the combined parameter $k_r/(C r_0^2)$. The spot size $r_0$ was pre-determined using the beam-offset method with a 2% uncertainty, and the heat capacity of glass was assumed to be $C = 1.65$ MJ/(m$^3 \cdot$K) with a 3% uncertainty. By best fitting the signals in Fig. 3(a1), the in-plane thermal conductivity of the glass plate was determined to be $k_r = 0.79$ W/(m·K), with an estimated uncertainty of 5%.

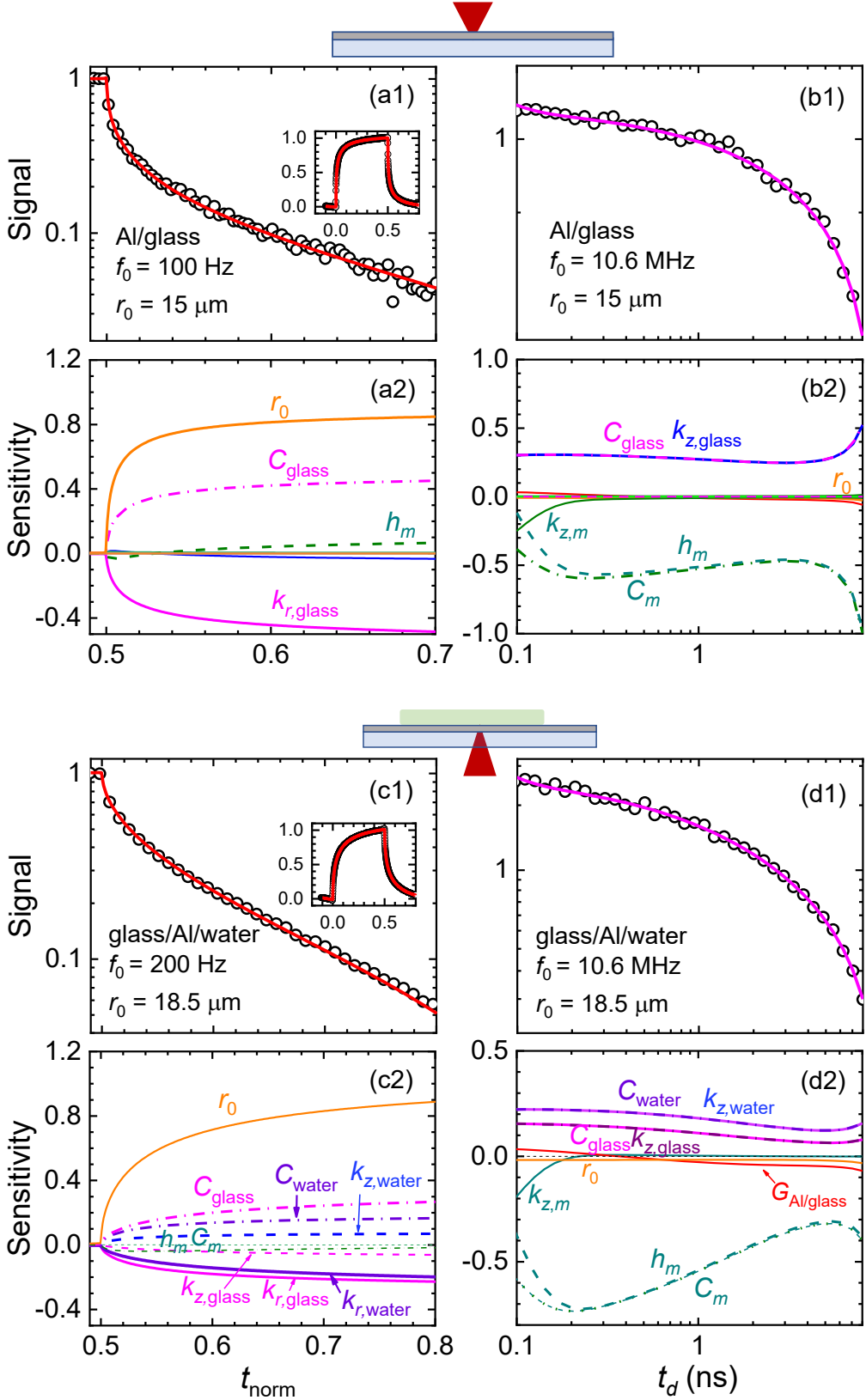

**FIG. 3** Measured signals and sensitivity analysis for Al/glass and glass/Al/water samples using PWA-TDTR and conventional TDTR methods. (a) Al/glass measured with PWA-TDTR at 100 Hz; (b) Al/glass measured with conventional TDTR at 10.6 MHz; (c) Glass/Al/water measured with PWA-TDTR at 200 Hz; (d) Glass/Al/water measured with conventional TDTR at 10.6 MHz. (a1-d1) Corresponding measured signals; (a2-d2) Sensitivity analysis.

Figure 3(b1) shows the signals measured on the same sample using the conventional TDTR method with the same spot size at a modulation frequency of 10.6 MHz. Sensitivity analysis in Fig. 3(b2) reveals that this signal set is primarily sensitive to $\sqrt{k_z C}/(h_m C_m)$. Assuming an isotropic thermal conductivity for the glass plate and using the literature value for the heat capacity of Al ($C_m = 2.44 \pm 0.07$ MJ/(m$^3 \cdot$ K) [16]), we determine the Al film thickness to be $h_m = 77$ nm, with an estimated uncertainty of 4.2%.

This highlights a key advantage of combining conventional TDTR with PWA-TDTR. In the past, when only high-frequency TDTR data were available, multiple unknown parameters (such as $k_z$ and $h_m$) were coupled, necessitating the pre-determination of one parameter through other techniques to solve for the other in the thermal model. Typically, $h_m$ could be determined from picosecond acoustics inherent in TDTR signals, with a typical uncertainty of 5%. However, this method was ineffective in the present case because the acoustic echo from the Al/glass interface was too weak to be detectable. By incorporating PWA-TDTR, both the $k_z$ and $h_m$ can be simultaneously determined with reduced uncertainty, effectively overcoming this limitation.

We then perform measurements with the liquid applied to the surface of the transducer layer. Both the pump and probe beams pass through the transparent substrate and are focused on the Al film at the glass/Al interface. In this case, a bidirecitonal heat transfer model is used to simulate the signals. The mathematical expressions for the model remain the same as those in Section 2.2, with the only modification being in the Green's function $\hat{G}(\rho, \omega)$ in Eq. (3). Further details on the Green's function for the bidirecitonal heat transfer case can be found in Ref. [21] and Supplementary Information Section S1.

Figure 3(c1) shows the signals measured on the glass/Al/water structure using PWA-TDTR with a spot size of $r_0 = 18.5$ μm and a modulation frequency of 200 Hz, while Fig. 3(d1) shows the conventional TDTR signals at a modulation frequency of 10.6 MHz. Sensitivity analysis in Fig. 3(c2) and 3(d2) indicates that the PWA-TDTR signals are primarily sensitive to $k_{r,\text{glass}}/(C_{\text{glass}} r_0^2)$ and $k_{r,\text{water}}/(C_{\text{water}} r_0^2)$, with some slight sensitivity to $\sqrt{k_{z,\text{water}} C_{\text{water}}}/(h_m C_m)$ and $\sqrt{k_{z,\text{glass}} C_{\text{glass}}}/(h_m C_m)$. In contrast, the TDTR signals are primarily sensitive to $\sqrt{k_{z,\text{water}} C_{\text{water}}}/(h_m C_m)$ and $\sqrt{k_{z,\text{glass}} C_{\text{glass}}}/(h_m C_m)$, with some additional sensitivity to $G_{\text{Al/glass}}/(h_m C_m)$.

With the properties of the Al film ($h_m C_m$), glass substrate ($k_{\text{glass}}$, $C_{\text{glass}}$), and the spot size ($r_0$) pre-determined, best-fitting the signals in Fig. 3(c1) and 3(d1) yields $k_{\text{water}} = 0.605 \pm 0.035$ W/(m·K), $C_{\text{water}} = 4.18 \pm 0.24$ MJ/(m³·K), both with an uncertainty of approximately 7%. These values agree well with established literature data [26, 27]. Additionally, the interfacial thermal conductance between Al and water is determined to be $G_{\text{Al/glass}} = 61 \pm 15$ MW/(m²·K), corresponding to a Kapitza length of ~10 nm, which is consistent with previous experimental studies [23]. This approach can also be extended to measure other liquids and soft gels.

## C. Probing depth-dependent thermal conductivity in LiNbO$_3$

The extended frequency coverage of PWA-TDTR allows for probing thermal conductivity profiles at depths up to hundreds of micrometers. This capability is particularly valuable for materials like lithium niobate (LiNbO$_3$), a critical component in nonlinear optics and electro-optic devices, where effective thermal management is essential for device performance and stability. In high-power applications, efficient heat dissipation reduces temperature-induced drift, while point defects, such as oxygen vacancies (VO), can significantly reduce thermal conductivity and degrade device performance.

In this work, we employ a combination of PWA-TDTR and conventional TDTR to investigate how depth-dependent distributions of VO affect the thermal conductivity of LiNbO$_3$. The VO profile is introduced by thermally reducing an X-cut LiNbO$_3$ crystal in an argon-hydrogen atmosphere. During this process, hydrogen atoms preferentially react with oxygen atoms bonded to lithium, breaking Li-O bonds and creating oxygen vacancies, along with a LiNb$_3$O$_8$ surface layer. As the annealing temperature increases to 800°C, more Li-O bonds are disrupted, generating additional oxygen vacancies and some lithium vacancies. These defects induce changes in crystal color, reduce the oxidation state of Nb ions from $Nb^{5+}$ to $Nb^{4+}$, and lead to local structural rearrangements and the formation of polarons and new phases.

To capture the effect of the defect distribution on thermal conductivity, we model the depth-dependent thermal conductivity of the reduced $LiNbO_3$ crystal using an exponential decay profile:

$$k(z) = k_{bulk} + (k_0 - k_{bulk})e^{-\frac{z}{\lambda}} \tag{9}$$

where $k_0$ is the surface thermal conductivity, $k_{bulk}$ is the thermal conductivity at sufficient depth, and $\lambda$ is the decay constant that characterize the depth at which $k(z)$ falls to $e^{-1}$ of its surface deviation. The parameters $k_0$, $k_{bulk}$, and $\lambda$ are extracted by fitting the experimental data from both conventional TDTR and PWA-TDTR measurements.

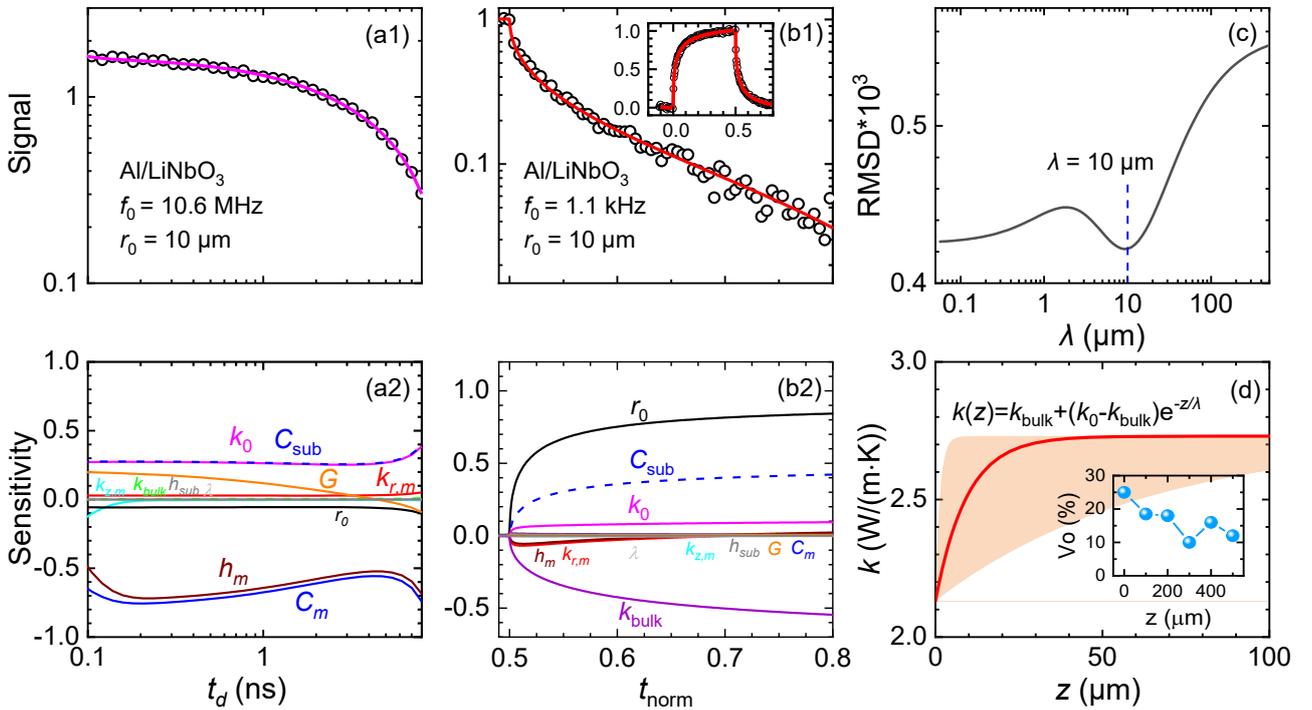

**FIG. 4.** Measurement Results for the $LiNbO_3$ sample with non-uniform defects. (a1, b1) Comparison of signals and model fits for convecntional TDTR measurement at 10.6 MHz and PWA-TDTR measurement at 1.1 kHz, respectively. Symbols represent the measured signals, and curves represent the model predictions; (a2, b2) Sensitivity curves of the signals to various parameters at high and low frequencies, respectively; (c) Calculated root-mean-square deviation (RMSD) between experimental data and model predictions as a function of the decay constant $\lambda$. The optimal $\lambda$ value is determined by minimizing the RMSD; (d) Depth-dependent variation of $k$ for the sample. The shaded regions represent the measurement uncertainty. Inset shows depth profiles of VO concentration determined by XPS analysis.

Figure 4 shows the measurement results, where Fig. 4(a1) shows conventional TDTR measurements using a laser spot size of $r_0 = 10$ μm and a high modulation frequency of $f_0 =$ 10.6 MHz, while Fig. 4(b1) displays PWA-TDTR measurements using the same spot size but at a reduced modulation frequency of $f_0 = 1.1$ kHz. The corresponding sensitivity coefficients are illustrated in Fig. 4(a2) and 4(b2).

Sensitivity analysis reveals distinct parameter dependencies. At 10.6 MHz, conventional TDTR signals are primarily sensitive to the combined parameters $\frac{\sqrt{k_0 C}}{h_m C_m}$ and $\frac{G}{h_m C_m}$. In contrast, at 1.1 kHz, PWA-TDTR signals are mainly sensitive to the combined parameter $\frac{k_{\text{bulk}}}{Cr_0^2}$. Given known values for $h_m C_m$, $C$, and $r_0$, we determine the surface thermal conductivity as $k_0 = 2.15 \pm 0.15$ W/(m·K), the bulk thermal conductivity as $k_{\text{bulk}} = 2.73 \pm 0.16$ W/(m·K), and the interfacial thermal conductance as $G_{\text{Al/LiNbO3}} = 36 \pm 3$ MW/(m²·K) by jointly fitting both datasets.

Since the signals exhibit low sensitivity to the decay constant $\lambda$, we varied its value to optimize the fit for both datasets. The fitting quality, quantified by the root-mean-square deviation (RMSD) between the experimental data and the simulated signals and plotted in Fig. 4(c) as a function of $\lambda$, has an optimal value when $\lambda$ is 10 μm. Given a non-symmetric distribution of the possible values of $\lambda$, we applied a logarithmic transformation (ln($\lambda$)) to approximate a normal distribution. Using full error propagation, we calculated the error of ln($\lambda$), from which we determine the error range for $\lambda$ as $\lambda = 10^{+18}_{-8}$ μm.

Figure 4(d) shows the $k$ depth profile, where $k$ increases from 2.1 to 2.8 W/(m·K) within the first ~50 μm before stabilizing. This depth dependence is attributed to the non-uniform distribution of VO, a dominant point defect in the sample. The inset of Fig. 4(d) illustrates the VO concentration decreasing with depth, explaining the corresponding increase in $k$. This behavior is consistent with the thermal reduction mechanism of LiNbO$_3$ in a hydrogen atmosphere, where reduction initiates at the surface and weakens toward the interior, resulting in the highest VO concentration at the surface and a gradual decrease with depth.

## IV. CONCLUSIONS

In summary, we have introduced PWA-TDTR, a novel technique that extends the low-frequency limit of conventional TDTR down to 50 Hz. By leveraging the pulse accumulation effect and conducting periodic waveform analysis of temperature response signals from square-wave modulated laser heating, PWA-TDTR overcomes the limitations of conventional TDTR, enabling accurate measurements of in-plane thermal conductivities as low as 0.2 W/(m·K). The method integrates easily with existing TDTR setups with minimal modifications, providing a practical and versatile solution for thermal property characterization at low modulation frequencies.

Validation experiments on PMMA and silica substrates confirmed the robustness of PWA-TDTR, with thermal conductivity and heat capacity values in excellent agreement with literature data. The technique has also been successfully applied to measure the thermal conductivity and heat capacity of thin liquid films and to probe depth-dependent thermal conductivity variations in lithium niobate crystals, revealing defect-induced inhomogeneities over a 100 μm depth. These applications highlight the potential of PWA-TDTR for investigating complex thermal transport phenomena in a wide range of materials, including soft matter, liquids, and anisotropic crystals.

By bridging the gap between high-frequency TDTR and low-frequency thermoreflectance techniques, PWA-TDTR significantly expands the measurable range for thermal property measurements. Its ability to simultaneously determine both cross-plane and in-plane thermal properties with high sensitivity and accuracy positions it as a powerful tool for advancing research in fields such as electronics, photonics, and thermal management.

## SUPPLEMENTARY MATERIAL

See the supplementary material for a detailed derivation of the thermal model used in the PWA-TDTR technique.

## ACKNOWLEDGMENTS

This work was supported by the National Key Research and Development Program of China (2022YFB3803900) and the National Natural Science Foundation of China (NSFC) through Grant No. 52376058.

## AUTHOR DECLARATIONS

**Conflict of Interest**

The authors have no known competing interest to declare.

## DATA AVAILABILITY

The data that support the findings of this study are available from the corresponding author upon reasonable request.

## REFERENCES

1. Cahill, D. G. *MRS Bull.* **2018,** 43, (10), 782-789.
2. Chiritescu, C.; Cahill, D. G.; Nguyen, N.; Johnson, D.; Bodapati, A.; Keblinski, P.; Zschack, P. *Science* **2007,** 315, (5810), 351-353.
3. Wang, Z.; Carter, J. A.; Lagutchev, A.; Koh, Y. K.; Seong, N.-H.; Cahill, D. G.; Dlott, D. D. *Science* **2007,** 317, (5839), 787.
4. Chen, K.; Song, B.; Ravichandran, N. K.; Zheng, Q.; Chen, X.; Lee, H.; Sun, H.; Li, S.; Gamage, G. A.; Tian, F.; Ding, Z.; Song, Q.; Rai, A.; Wu, H.; Koirala, P.; Schmidt, A. J.; Watanabe, K.; Lv, B.; Ren, Z.; Shi, L.; Cahill, D. G.; Taniguchi, T.; Broido, D.; Chen, G. *Science* **2020,** eaaz6149.
5. Costescu, R. M.; Cahill, D. G.; Fabreguette, F. H.; Sechrist, Z. A.; George, S. M. *Science* **2004,** 303, (5660), 989-990.
6. Kang, K.; Koh, Y. K.; Chiritescu, C.; Zheng, X.; Cahill, D. G. *Rev. Sci. Instrum.* **2008,** 79, (11), 114901.
7. Wei, C.; Zheng, X.; Cahill, D. G.; Zhao, J. C. *Rev. Sci. Instrum.* **2013,** 84, (7), 071301.
8. Larkin, L. S.; Smoyer, J. L.; Norris, P. M. *Int. J. Heat Mass Transfer* **2017,** 109, 786-790.
9. Chen, T.; Song, S.; Shen, Y.; Zhang, K.; Jiang, P. *International Communications in Heat and Mass Transfer* **2024,** 158, 107849.
10. Jiang, P.; Qian, X.; Yang, R. *J. Appl. Phys.* **2018,** 124, (16), 161103.
11. Chen, T.; Song, S.; Hu, R.; Jiang, P. *International Journal of Thermal Sciences* **2025,** 207, 109347.
12. Song, S.; Chen, T.; Jiang, P. *J. Appl. Phys.* **2025,** 137, (5), 055101.
13. Yang, J.; Ziade, E.; Schmidt, A. J. *Rev. Sci. Instrum.* **2016,** 87, (1), 014901.
14. Chen, T.; Jiang, P.-Q. *Acta Physica Sinica* **2024,** 73, (23), 230202.
15. Chen, T.; Jiang, P. *Physical Review Applied* **2025,** 23, (4), 044004.
16. Touloukian, Y.; Buyco, E. *Thermophysical Properties of Matter-The TPRC Data Series. Volume 4. Specific Heat-Metallic Elements and Alloys*; DTIC Document: 1971.
17. Assael, M. J.; Botsios, S.; Gialou, K.; Metaxa, I. N. *Int. J. Thermophys.* **2005,** 26, (5), 1595-1605.
18. Assael, M. J.; Antoniadis, K. D.; Wu, J. *Int. J. Thermophys.* **2008,** 29, (4), 1257-1266.
19. Brückner, R. *J. Non·Cryst. Solids* **1970,** 5, (2), 123-175.
20. Sugawara, A. *J. Appl. Phys.* **1968,** 39, (13), 5994-5997.
21. Schmidt, A.; Chiesa, M.; Chen, X.; Chen, G. *Rev. Sci. Instrum.* **2008,** 79, (6), 064902.
22. Sun, F.; Zhang, T.; Jobbins, M. M.; Guo, Z.; Zhang, X.; Zheng, Z.; Tang, D.; Ptasinska, S.; Luo, T. *Adv. Mater.* **2014,** 26, (35), 6093-6099.
23. Ge, Z.; Cahill, D.; Braun, P. *Phys. Rev. Lett.* **2006,** 96, (18), 186101.
24. Tian, Z.; Marconnet, A.; Chen, G. *Appl. Phys. Lett.* **2015,** 106, (21), 211602.
25. Cho, J.; Losego, M. D.; Zhang, H. G.; Kim, H.; Zuo, J.; Petrov, I.; Cahill, D. G.; Braun, P. V. *Nat. Commun.* **2014,** 5, 4035.
26. Ramires, M. L. V.; Nieto de Castro, C. A.; Nagasaka, Y.; Nagashima, A.; Assael, M. J.; Wakeham, W. A. *J. Phys. Chem. Ref. Data* **1995,** 24, (3), 1377-1381.

27. Mallamace, F.; Corsaro, C.; Mallamace, D.; Fazio, E.; Chen, S.-H.; Cupane, A. *International Journal of Molecular Sciences* **2020,** 21, (2), 622.

# Extending the Low-Frequency Limit of Time-Domain Thermoreflectance via Periodic Waveform Analysis

Mingzhen Zhang[1], Tao Chen[1], Shangzhi Song[1], Yunjia Bao[2,3], Ruiqiang Guo[3], Weidong Zheng[3,*], Puqing Jiang[1,*], Ronggui Yang[1,4,*]

[1]School of Energy and Power Engineering, Huazhong University of Science and Technology, Wuhan, Hubei 430074, China

[2]Institute of Novel Semiconductors, State Key Laboratory of Crystal Materials, Shandong University, Jinan, Shandong 250100, China

[3]Thermal Science Research Center, Shandong Institute of Advanced Technology, Jinan, Shandong 250103, China

[4]Department of Energy and Resource Engineering, College of Engineering, Peking University, Beijing 100871, China

## S1. Modeling of Signals Acquired in PWA-TDTR Experiments

This section provides a detailed derivation of the signals obtained in periodic waveform analysis-based time-domain thermoreflectance (PWA-TDTR) experiments.

*1. Pump Beam Intensity in Time and Space Domains*

Assuming the laser pulses can be approximated by a Dirac delta function—a valid assumption given the ultrashort pulse duration (<500 fs) relative to the inter-pulse interval (~12.5 ns)—the pump beam intensity in the real space and time domain is expressed as:

$$p_1(r,t) = \frac{2A_1}{\pi r_1^2} \exp\left(-\frac{2r^2}{r_1^2}\right) \sum_{m=0}^{\infty} \frac{4}{i(2m+1)\pi} e^{i(2m+1)\omega_0 t} \sum_{n=-\infty}^{\infty} \delta(t - nT_s - t_0) \qquad \text{(S1-1)}$$

This represents a train of delta functions modulated by a 50% duty cycle square wave at frequency $\omega_0$. Here, $f_{rep}$ is the laser repetition frequency with period $T_s = 1/f_{rep} = 2\pi/\omega_s$, $A_1$ is the average pump power, spatially distributed as a Gaussian with a $1/e^2$ radius $r_1$, and $t_0$ is an arbitrary time shift of the laser pulses.

The complex exponential $e^{i\omega_0 t}$ compactly represents oscillatory behavior, encoding both cosine ($\cos(\omega_0 t)$) and sine ($\sin(\omega_0 t)$) components of the signal. This form simplifies frequency-domain analysis, facilitating operations such as modulation, mixing, and filtering. The real part corresponds to the measurable physical field, while the complex representation aids in phase and frequency analysis.

2. *Frequency-Domain Representation of the Pump Beam*

The Fourier transform of the delta train yields a delta train in the angular frequency domain with spacing $\omega_s = \frac{2\pi}{T_s}$:

$$S = \int_{-\infty}^{\infty} \sum_{n=-\infty}^{\infty} \delta(t - nT_s - t_0) e^{-i\omega t} dt$$

$$= e^{-i\omega t_0} \sum_{n=-\infty}^{\infty} e^{-in\omega T_s}$$

$$= e^{-i\omega t_0} \omega_s \sum_{n=-\infty}^{\infty} \delta(\omega - n\omega_s)$$

$$= \omega_s \sum_{n=-\infty}^{\infty} \delta(\omega - n\omega_s) e^{-in\omega_s t_0}$$

The Fourier transform of the square wave is

$$F = \int_{-\infty}^{\infty} \sum_{m=0}^{\infty} \frac{4}{i(2m+1)\pi} e^{i(2m+1)\omega_0 t} e^{-i\omega t} dt$$

$$= \sum_{m=0}^{\infty} \frac{4}{i(2m+1)\pi} \int_{-\infty}^{\infty} e^{i(2m+1)\omega_0 t} e^{-i\omega t} dt$$

$$= \sum_{m=0}^{\infty} \frac{4}{i(2m+1)\pi} 2\pi\delta(\omega - (2m+1)\omega_0) \tag{S1-2}$$

The product in the time domain becomes a convolution in the frequency domain. Thus, the Fourier transform of $s(t)f(t)$ is:

$$H = \int_{-\infty}^{\infty} s(t)f(t)e^{-i\omega t} dt = \frac{1}{2\pi} \int_{-\infty}^{\infty} S(\theta)F(\omega - \theta) d\theta$$

$$= \frac{1}{2\pi} \int_{-\infty}^{\infty} \omega_s \sum_{n=-\infty}^{\infty} \delta(\theta - n\omega_s) e^{-in\omega_s t_0} \sum_{m=0}^{\infty} \frac{4}{i(2m+1)\pi} 2\pi\delta(\omega - \theta - (2m+1)\omega_0) d\theta$$

$$= \int_{-\infty}^{\infty} \omega_s \sum_{n=-\infty}^{\infty} \delta(\theta - n\omega_s) e^{-in\omega_s t_0} \sum_{m=0}^{\infty} \frac{4}{i(2m+1)\pi} \delta(\omega - \theta - (2m+1)\omega_0) d\theta$$

$$= \omega_s \sum_{n=-\infty}^{\infty} e^{-in\omega_s t_0} \sum_{m=0}^{\infty} \frac{4}{i(2m+1)\pi} \delta(\omega - n\omega_s - (2m+1)\omega_0)$$

Incorporating the Hankel transform over $r$, the frequency-domain pump power expression becomes:

$$P_1(\rho, \omega) = A_1 e^{-\pi^2 \rho^2 r_1^2/2} \omega_s \sum_{n=-\infty}^{\infty} \sum_{m=0}^{\infty} \frac{4}{i(2m+1)\pi} \delta(\omega - n\omega_s - (2m+1)\omega_0) e^{-in\omega_s t_0}$$

(S1-3)

For comparison, in conventional TDTR:

$$P_1(\rho, \omega) = A_1 e^{-\pi^2 \rho^2 r_1^2/2} \omega_s \sum_{n=-\infty}^{\infty} \delta(\omega - n\omega_s - \omega_0) e^{-in\omega_s t_0} \tag{S1-3'}$$

3. *Surface Temperature Response*

The surface temperature response in the frequency domain is the product of the heat input $P_1$ and the system's thermal response function $\hat{G}$:

$$\Theta(\rho, \omega) = P_1(\rho, \omega) \hat{G}(\rho, \omega) \tag{S1-4}$$

The inverse Hankel transform yields the surface temperature distribution:

$$\Theta(r,\omega) = \int_0^\infty P_1(k,\omega)\hat{G}(k,\omega)J_0(2\pi kr)2\pi k dk \tag{S1-5}$$

The inverse Fourier transform of $\Theta(r,\omega)$ provides the time-domain surface temperature response $\theta(r,t)$.

4. *Probe Beam and Signal Detection*

The probe beam, delayed by time $t_d$ relative to the pump, is expressed as:

$$p_2(r,t) = \frac{2A_2}{\pi r_2^2}\exp\left(-\frac{2r^2}{r_2^2}\right)\sum_{m=-\infty}^{\infty}\delta(t - mT_s - t_0 - t_d) \tag{S1-6}$$

Its Fourier transform is:

$$P_2(r,\omega) = \frac{2A_2}{\pi r_2^2}\exp\left(-\frac{2r^2}{r_2^2}\right)\sum_{m=-\infty}^{\infty}\delta(\omega - m\omega_s)e^{-im\omega_s(t_0+t_d)} \tag{S1-7}$$

The probe samples a weighted average of the temperature distribution, given by the convolution:

$$\Delta\Theta(\omega) = \int_0^\infty \left(\frac{1}{2\pi}\int_{-\infty}^{\infty}\Theta(r,\zeta)P_2(r,\omega-\zeta)d\zeta\right)2\pi r dr \tag{S1-8}$$

That is:

$$\Delta\Theta(\omega) =$$

$$\int_0^\infty\left(\frac{1}{2\pi}\int_{-\infty}^\infty \Theta(r,\zeta)\frac{2A_2}{\pi r_2^2}\exp\left(-\frac{2r^2}{r_2^2}\right)\sum_{m=-\infty}^\infty \delta(\omega-\zeta-m\omega_s)e^{-im\omega_s(t_0+t_d)}\,d\zeta\right)2\pi r dr$$

$$= \int_0^\infty\left(\frac{1}{2\pi}\int_{-\infty}^\infty\left(\int_0^\infty \Theta(\rho,\zeta)J_0(2\pi\rho r)\,2\pi\rho d\rho\right)\sum_{m=-\infty}^\infty\delta(\omega-\zeta$$

$$- m\omega_s)e^{-im\omega_s(t_0+t_d)}\,d\zeta\right)\frac{2A_2}{\pi r_2^2}\exp\left(-\frac{2r^2}{r_2^2}\right)2\pi r dr$$

$$= \int_0^\infty \left( \frac{1}{2\pi} \int_{-\infty}^\infty \Theta(\rho,\zeta) \left( \int_0^\infty \frac{2A_2}{\pi r_2^2} \exp\left(-\frac{2r^2}{r_2^2}\right) J_0(2\pi\rho r) 2\pi r dr \right) \sum_{m=-\infty}^\infty \delta(\omega - \zeta$$
$$- m\omega_s) e^{-im\omega_s(t_0+t_d)} \, d\zeta \right) 2\pi\rho d\rho$$

$$= \frac{1}{2\pi} \int_0^\infty \left( \int_{-\infty}^\infty \Theta(\rho,\zeta) \sum_{m=-\infty}^\infty \delta(\omega - \zeta - m\omega_s) e^{-im\omega_s(t_0+t_d)} \, d\zeta \right) A_2 e^{-\pi^2\rho^2 r_2^2/2} 2\pi\rho d\rho$$

$$= \frac{1}{2\pi} \int_0^\infty \left( \int_{-\infty}^\infty P_1(\rho,\zeta) \hat{G}(\rho,\zeta) \sum_{m=-\infty}^\infty \delta(\omega - \zeta - m\omega_s) e^{-im\omega_s(t_0+t_d)} \, d\zeta \right) A_2 e^{-\pi^2\rho^2 r_2^2/2} 2\pi\rho d\rho$$

$$= \frac{1}{2\pi} \int_0^\infty \left( \int_{-\infty}^\infty A_1 e^{-\pi^2\rho^2 r_1^2/2} \omega_s \sum_{n=-\infty}^\infty \sum_{m_1=0}^\infty \frac{4}{i(2m_1+1)\pi} \delta(\zeta - n\omega_s \right.$$
$$- (2m_1+1)\omega_0) e^{-in\omega_s t_0} \hat{G}(\rho,\zeta) \sum_{m=-\infty}^\infty \delta(\omega - \zeta$$
$$\left. - m\omega_s) e^{-im\omega_s(t_0+t_d)} \, d\zeta \right) A_2 e^{-\pi^2\rho^2 r_2^2/2} 2\pi\rho d\rho$$

$$= \frac{\omega_s}{2\pi} \int_0^\infty \left( \sum_{n=-\infty}^\infty \sum_{m_1=0}^\infty \frac{4}{i(2m_1+1)\pi} e^{-in\omega_s t_0} \hat{G}(\rho, n\omega_s \right.$$
$$+ (2m_1+1)\omega_0) \sum_{m=-\infty}^\infty \delta(\omega - n\omega_s - (2m_1+1)\omega_0$$
$$\left. - m\omega_s) e^{-im\omega_s(t_0+t_d)} \right) A_1 A_2 e^{-\pi^2\rho^2 r_0^2} 2\pi\rho d\rho$$

(S1-9)

The delta function evaluates to zero at all frequencies except at $\omega = (n+m)\omega_s - (2m_1+1)\omega_0$, where $m$, $n$, and $m_1$ are integers. The lock-in makes sure the detected frequency is around the reference frequency. Since $\omega_s$ is at least one order greater than $\omega_0$, we see that the above term is non-zero only when $m = -n$. Thus, Eq. (S1-9) could be simplified as:

$$\Delta\Theta(\omega) = \frac{\omega_s}{2\pi} A_1 A_2 \int_0^\infty \left( \sum_{n=-\infty}^{\infty} \sum_{m_1=0}^{\infty} \frac{4}{i(2m_1+1)\pi} \delta(\omega - (2m_1+1)\omega_0) \hat{G}(\rho, n\omega_s) \right.$$

$$\left. + (2m_1+1)\omega_0) e^{in\omega_s t_d} \right) e^{-\pi^2 \rho^2 r_0^2} 2\pi\rho d\rho$$

(S1-10)

where $r_0 = \sqrt{(r_1^2 + r_2^2)/2}$ is the root mean square (RMS) average of the pump and probe $1/e^2$ radii.

For comparison, in conventional TDTR, the same signal is:

$$\Delta\Theta(\omega) = A_1 \int_0^\infty \sum_{n=-\infty}^{\infty} \delta(\omega - \omega_0) \hat{G}(k, \omega_0 + n\omega_s) e^{in\omega_s t_d} \exp(-\pi^2 k^2 r_0^2) 2\pi k dk \quad \text{(S1-10')}$$

If we define the temperature response due to the harmonic heating at frequency $\omega$ as

$$\Delta T(\omega) = \frac{\omega_s}{2\pi} A_1 A_2 \int_0^\infty \hat{G}(\rho, \omega) \exp(-\pi^2 \rho^2 w_0^2) 2\pi\rho d\rho \quad \text{(S1-11)}$$

Eq. (S1-10) can then be re-written as

$$\Delta\Theta(\omega) = \sum_{n=-\infty}^{\infty} \sum_{m_1=0}^{\infty} \frac{4}{i(2m_1+1)\pi} \delta(\omega - (2m_1+1)\omega_0) \Delta T(n\omega_s + (2m_1+1)\omega_0) e^{in\omega_s t_d}$$

$$= \sum_{m_1=0}^{\infty} \frac{4}{i(2m_1+1)\pi} \delta(\omega - (2m_1+1)\omega_0) \sum_{n=-\infty}^{\infty} \Delta T(n\omega_s + (2m_1+1)\omega_0) e^{in\omega_s t_d}$$

(S1-12)

For comparison, in conventional TDTR, the same signal is:

$$\Delta\Theta(\omega) = \delta(\omega - \omega_0) \sum_{n=-\infty}^{\infty} \Delta T(n\omega_s + \omega_0) e^{in\omega_s t_d} \quad \text{(S1-12')}$$

Inverse Fourier transform of Eq. (S1-12) gives the probed signal in the time domain at delay time $t_d$ as:

$$\Delta\Theta(t) = \sum_{m_1=0}^{\infty} \frac{4}{i(2m_1+1)\pi} e^{i(2m_1+1)\omega_0 t} \sum_{n=-\infty}^{\infty} \Delta T(n\omega_s + (2m_1+1)\omega_0) e^{in\omega_s t_d} \quad \text{(S1-13)}$$

For comparison, in conventional TDTR, the same signal is:

$$\Delta T(t) = e^{i\omega_0 t} \sum_{n=-\infty}^{\infty} \Delta T(n\omega_s + \omega_0) e^{in\omega_s t_d} \quad \text{(S1-13')}$$

## S2. Derivation of the Green's Function for a Multilayered System

The Green's function is derived by solving the heat diffusion equation in multilayered systems:

$$C \frac{\partial T}{\partial t} = \frac{k_r}{r} \frac{\partial}{\partial r} \left( r \frac{\partial T}{\partial r} \right) + k_z \frac{\partial^2 T}{\partial z^2} \quad \text{(S2-1)}$$

where $k_r$ and $k_z$ are the thermal conductivities of the sample in the radial and through-plane directions, respectively, and $C$ is the volumetric heat capacity.

In the frequency domain, this simplifies to:

$$\frac{\partial^2 \Theta}{\partial z^2} = \lambda^2 \Theta \quad \text{(S2-2)}$$

where $\lambda^2 = (4\pi^2 \rho^2 k_r + i\omega C)/k_z$, and $\rho$ is the transform variable introduced by the Hankel transform.

The general solution of Eq. (S2-2) is

$$\Theta = e^{\lambda z} B^+ + e^{-\lambda z} B^- \quad \text{(S2-3)}$$

with $B^+, B^-$ determined by boundary conditions.

The heat flux can be obtained from the Fourier's law of heat conduction $Q = -k_z (d\Theta/dz)$ and Eq. (S2-3). The temperature and heat flux in the $n$-th layer at depth $z = L$ from the surface of that layer can be expressed in matrix form as:

$$\begin{bmatrix} \Theta \\ Q \end{bmatrix}_{n, z=L} = [N]_n \begin{bmatrix} B^+ \\ B^- \end{bmatrix}_n \quad \text{(S2-4)}$$

where $[N]_n$ is defined as:

$$[N]_n = \begin{bmatrix} 1 & 1 \\ -\gamma_n & \gamma_n \end{bmatrix} \begin{bmatrix} e^{\lambda L} & 0 \\ 0 & e^{-\lambda L} \end{bmatrix}_n \quad \text{(S2-5)}$$

where $\gamma = k_z \lambda$.

The constants $B^+, B^-$ can thus be viewed as the properties of the $n$th layer, which can also be obtained from the surface temperature and heat flux of that layer as

$$\begin{bmatrix} B^+ \\ B^- \end{bmatrix}_n = [M]_n \begin{bmatrix} \Theta \\ Q \end{bmatrix}_{n,z=0} \tag{S2-6}$$

where $[M]_n$ is defined as:

$$[M]_n = \frac{1}{2\gamma_n} \begin{bmatrix} \gamma_n & -1 \\ \gamma_n & 1 \end{bmatrix} \tag{S2-7}$$

For heat flow across interfaces, there is another matrix to relate the temperature and heat flux at the bottom of the upper layer to those at the top of the underlayer as

$$\begin{bmatrix} \Theta \\ Q \end{bmatrix}_{n+1,z=0} = [R]_n \begin{bmatrix} \Theta \\ Q \end{bmatrix}_{n,z=L} \tag{S2-8}$$

where $[R]_n$ is defined as:

$$[R]_n = \begin{bmatrix} 1 & -1/G_n \\ 0 & 1 \end{bmatrix} \tag{S2-9}$$

where $G_n$ is the interface conductance between the $n$th and $(n+1)$th layers. The temperature and heat flux on the top surface of the multilayer stack can thus be related to those at the bottom of the substrate as:

$$\begin{bmatrix} \Theta \\ Q \end{bmatrix}_{n,z=L_n} = [N]_n [M]_n \ldots [R]_1 [N]_1 [M]_1 \begin{bmatrix} \Theta \\ Q \end{bmatrix}_{1,z=0} = \begin{bmatrix} A & B \\ C & D \end{bmatrix} \begin{bmatrix} \Theta \\ Q \end{bmatrix}_{1,z=0} \tag{S2-10}$$

Applying the boundary condition that at the bottom of the substrate $Q_{z\to\infty} = 0$ yields $0 = C\Theta_{1,z=0} + DQ_{1,z=0}$. The thermal response function $\hat{G}$, which is the detected temperature response under a unit heat flux, can thus be found out as

$$\hat{G}(k,\omega) = \frac{\Theta_{1,z=0}}{Q_{1,z=0}} = -\frac{D}{C} \tag{S2-11}$$

For the case of bidirectional heat flow, we have two multilayered systems, each with a semi-infinite substrate. We thus have

$$\begin{bmatrix} \Theta \\ Q \end{bmatrix}_{n1} = \begin{bmatrix} A_1 & B_1 \\ C_1 & D_1 \end{bmatrix} \begin{bmatrix} \Theta \\ Q \end{bmatrix}_{s1}, \quad \begin{bmatrix} \Theta \\ Q \end{bmatrix}_{n2} = \begin{bmatrix} A_2 & B_2 \\ C_2 & D_2 \end{bmatrix} \begin{bmatrix} \Theta \\ Q \end{bmatrix}_{s2}, \tag{S2-12}$$

and $C_1\Theta_{s1} + D_1 Q_{s1} = 0$, $C_2\Theta_{s2} + D_2\Theta_{s2} = 0$. Applying the boundary condition that $\Theta_s = \Theta_{s1} = \Theta_{s2}$, and $Q_s = Q_{s1} + Q_{s2}$, the thermal response function becomes:

$$\hat{G}(k,\omega) = \frac{\Theta_s}{Q_s} = -\frac{1}{\frac{C_1}{D_1}+\frac{C_2}{D_2}} = -\frac{D_1 D_2}{C_1 D_2 + C_2 D_1} \tag{S2-13}$$

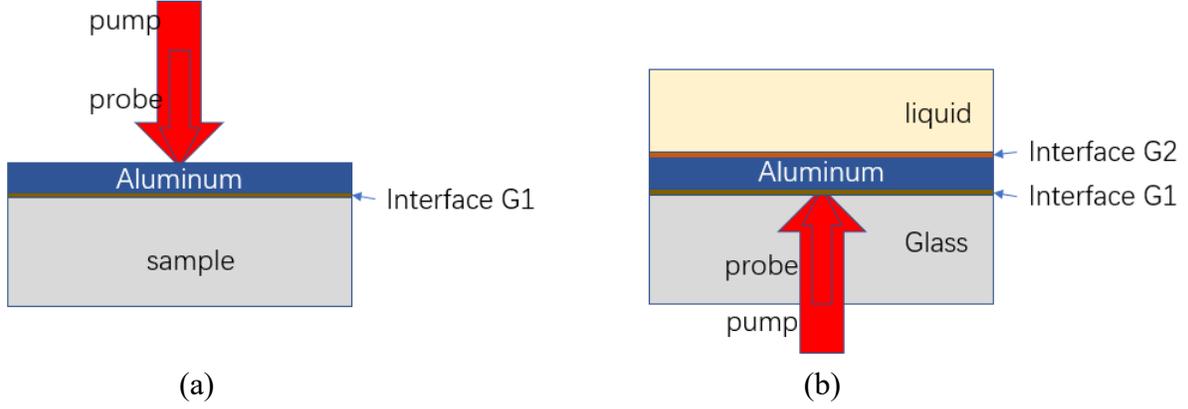

(a)           (b)

**FIG. S1.** Schematic diagrams of sample structures for (a) unidirectional heat transfer, where heating and probing are performed at the surface of multilayer structures, and (b) bidirectional heat transfer, where heating and probing occur at the interface between two semi-infinite substrates.

9